\documentclass[prl,twocolumn,superscriptaddress,showpacs,floatfix,amsfonts]{revtex4}
\usepackage{graphicx,graphics,color,epsfig}% Include figure files
\usepackage{bm}
\usepackage{amsmath}
\usepackage{amssymb}
\usepackage{epstopdf}
\begin{document}
\preprint{}
\title{The $f$-electron physics of rare-earth iron pnictides: influence of
$d$-electron antiferromagnetic order on the heavy fermion phase
diagram}
\author{Jianhui Dai}
\affiliation{Zhejiang Institute of Modern Physics, Zhejiang
University, Hangzhou 310027, China}
\author{Jian-Xin Zhu}
\affiliation{Theoretical Division, Los Alamos National Laboratory,
Los Alamos, New Mexico 87545, USA}
\author{Qimiao Si}
\affiliation{Department of Physics and Astronomy, Rice University,
Houston, TX 77005, USA}

\begin{abstract}
Some of the high $T_c$ iron pnictides contain magnetic
rare-earth elements, raising
the question of how the existence and
tunability of a $d$-electron antiferromagnetic order influences
the heavy fermion behavior of the $f$-moments. With CeOFeP
and CeOFeAs in mind as prototypes, we derive an extended
Anderson lattice model appropriate for these quaternary
systems. We show that the Kondo screening of the $f$-moments
are efficiently suppressed by the $d$-electron ordering.
We also argue that, inside the $d$-electron ordered
state (as in CeOFeAs), the $f$-moments provide a rare realization
of a quantum frustrated magnet with competing $J_1$-$J_2$-$J_3$
interactions in an effective square lattice. Implications
for the heavy fermion physics in broader contexts are also
discussed.
\end{abstract}
 \maketitle

\narrowtext

The homologous rare-earth iron arsenides exhibit antiferromagnetic
(AF) ground states in addition to the high temperature
superconductivity
\cite{Kamihara:JACS08,GFChen:08,Ren:08,XHChen:08,Wang:08}. The
systems of interest here are the arsenides $R$O$_x$F$_{1-x}$FeAs,
with $R$ = Ce, Sm, Nd, Pr,... being magnetic rare earths, which have
superconducting transition temperatures
higher~\cite{GFChen:08,Ren:08,XHChen:08,Wang:08} than the maximal
$T_c \approx 26\;\text{K}$ of
LaO$_x$F$_{1-x}$FeAs~\cite{Kamihara:JACS08}. The parent compounds of
these systems, ROFeAs, have a layered structure, with FeAs and $R$O
layers sandwiching each other. They typically show a collinear AF
order and a structure distortion, which are successively suppressed
by carrier doping in favor of superconductivity~\cite{Cruz:08}. Also
of interest are the iron phosphides. LaOFeP was the iron pnictide
reported to show superconductivity below $T_c\approx
4\;\text{K}$~\cite{Kamihara:06}.
%,McQueen:08,Hamlin:08}.
This compound
has the same layered structure as LaOFeAs, but does not order
magnetically~\cite{Kamihara:08}.

The distinction between the iron phosphides and arsenides becomes
even more pronounced when La is replaced by Ce. CeOFeP is neither
superconducting nor magnetically ordered, and its Ce $f$-electrons
exhibit heavy fermion behavior with a Kondo temperature $T_K\approx
10\;\text{K}$~\cite{Bruning:08}. CeOFeAs has the $d$-electron
collinear AF ordering below $T^{(d)}_N\approx 130\;\text{K}$. Its
$f$-electrons display a noticeable AF order below $T^{(f)}_N\approx
4\;\text{K}$~\cite{GFChen:08,Zhao:08}, but does not show any heavy
fermion features. What underlies the heavy fermion behavior in
CeOFeP and its absence in CeOFeAs? One possibility is that this
primarily reflects the very different interlayer $3d$-$4f$ couplings
between CeOFeP and CeOFeAs, as suggested by a first-principle
LDA+DMFT study~\cite{Georges:08}. However, a more complete theoretical
estimate using a full density of states, which is strongly peaked
away from the Fermi energy, suggests that the effective Kondo
couplings in CeOFeP and CeOFeAs may in fact be comparable~\cite{XDai:09}.
%A recent
%Experiments on muon spin relaxation and
Muon-spin-relaxation and neutron scattering
experiments \cite{Maeter:08,Chi:08}
%also indicates
may also be interpretted in terms of a sizable Kondo coupling in
CeOFeAs.

In this Communication, we discuss the possibility that the distinction in
the $d$-electron magnetism between CeOFeP and CeOFeAs plays an
important role in influencing their heavy fermion behavior. This
mechanism is expected to play an especially important role when we
consider not only the end materials CeOFeP and CeOFeAs, but also the
series CeOFeAs$_{1-x}$P$_{x}$, which has been proposed
to realize a continuously varying $d$-electron AF order and the
associated quantum critical point~\cite{DaiSiZhu:08}.

Studying the effect of the $d$-electron AF order on the heavy
fermion phase diagram not only sheds new light on the properties of
the iron pnictides, but also represents a new twist to the heavy
fermion physics in general. Typically, AF order in heavy fermion
metals is induced by the RKKY interactions among the $f$-moments,
and the heavy fermion phase diagram involves the competition between
RKKY and Kondo coupling~\cite{Gegenwart:08,Doniach:77}. A tunable
$d$-electron AF order adds a new dimension to the heavy fermion
phase diagram.

In the following, we will consider this effect within an extended
Anderson lattice model (ALM) appropriate for the stoichiometric
$R$-1111 compounds $R$OFe$X$ ($X$=As or P). The model incorporates
the inter-layer hybridization between pnictogen $X$ $p$-orbitals and
rare earth $R$ $f$-orbitals. We note in passing that the derived
model takes into account the microscopic crystal structure and
symmetry of the $R$-1111 compounds. Given that there are many
materials of the same ZrCuSiAs-type structure \cite{Rottgen:08},
with many of them containing magnetic rare-earth elements, we expect
that our model will also be germane to many such related compounds
\cite{Krellner:08}.

\begin{figure}[t]
\epsfxsize=8.0cm \centerline{\epsffile{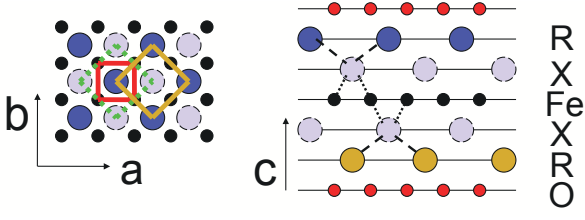}} \caption[]{(Color
online) The lattice structure of R-1111 series. The small black and
red (connected with solid line) circles represent Fe and O ions,
respectively, and the big blue/brown and dashed grey circles are the
$R$ and $X$ ions, respectively. The small solid red, solid brown,
and doted green squares describe the Fe, $R$, and $X$ plaquettes,
respectively. Left panel: $ab$-plane; right panel: $ac$-plane. The
dashed and doted lines denote $V_{pf}$ and $V_{pd}$, respectively.}
\label{fig:lattice}
\end{figure}

{\it General considerations.} The lattice structure of the $R$-1111
compound series is schematically shown in Fig.~\ref{fig:lattice}.
Let Fe-atoms be in the $(x,y)$-plane with the coordinate $({\vec
r},0)$, where, $\vec r=(i_x,i_y)$,  $i_x$ and $i_y$ are both
integers (the nearest Fe-Fe distance is set to unity). The
coordinates of $X$- and $R$-atoms are $({\vec r}_p,{\eta} z_p)$ and
$({\vec r}_f,\eta z_f)$ respectively, where ${\vec r}_p=(i_x
+1/2,i_y+1/2)$, ${\vec r}_f=(i_x-1/2,i_y+1/2)$, $\eta=e^{i\pi(i_x
+i_y)}$, $z_p$ and $z_f$ are the distances of $X$- and $R$-atoms to
the Fe-plane. We denote the $d$-, $p$-, and $f$-electrons by
$d^{(\alpha)}_{\sigma}({\vec r})$, $p^{(\mu)}_{\sigma}({\vec
r}_p,\eta z_p)$, and $f^{(m)}_{\sigma}({\vec r}_f,\eta z_f)$, with
orbital indices
$\alpha=d_{xy},\;d_{xz},\;d_{yz},\;d_{x^2-y^2},\;d_{3z^2-r^2}$,
$\mu=p_x,\; p_y,\; p_z$, and $m=1,\cdots,l$.

{\it The model Hamiltonian.} The hybridization part of the
Hamiltonian is given by $H_{hybrid}=H_{pd}+H_{pf}$, where
\begin{eqnarray}
H_{pd}=\sum_{{\vec r}}
V_{pd}^{(\mu,\alpha)}[p^{(\mu)\dagger}_{\sigma}({\vec r}_p,\eta
z_p)D^{(\alpha)}_{\sigma}({\vec r})
+\text{h.c.}]\;,~~~~~~~~\\
H_{pf}=\sum_{{\vec
r}}V_{pf}^{(\mu,m)}[p^{(\mu)\dagger}_{\sigma}({\vec r}_p, \eta
z_p)F^{(m)}_{\sigma}({\vec r}_f,\eta z_f) +\text{h.c.}]\;.
\end{eqnarray}
Here we introduce $D^{(\alpha)}_{\sigma}({\vec
r})=\sum_{\square}d^{(\alpha)}_{\sigma}({\vec r})\equiv
d^{(\alpha)}_{\sigma}(i_x,i_y)+d^{(\alpha)}_{\sigma}(i_x+1,i_y)+
d^{(\alpha)}_{\sigma}(i_x,i_y+1)+d^{(\alpha)}_{\sigma}(i_x+1,i_y+1)$
and $F^{(m)}_{\sigma}({\vec r}_f,\eta
z_f)=\sum_{\square}f^{(m)}_{\sigma}({\vec r}_f, \eta z_f)\equiv
f^{(m)}_{\sigma}(i_x-1/2, i_y+1/2,\eta
z_f)+f^{(m)}_{\sigma}(i_x+3/2, i_y+1/2,\eta z_f)+
f^{(m)}_{\sigma}(i_x+1/2, i_y-1/2,\eta
z_f)+f^{(m)}_{\sigma}(i_x+1/2, i_y+3/2,\eta z_f)$ as the plaquette
operators of $d$- and $f$-electrons around $X$-atoms. (Summations
over the repeated spin and channel indices are implied hereafter
unless otherwise specified.)

The interaction part of the Hamiltonian,
$H_{int}=H_{int,d}+H_{int,p}+H_{int,f}$, contains the usual on-site
Coulomb interactions ($U_p$, $U_d$, and $U_f$) and the Hund's
coupling ($J_{H,d}$). The total Hamiltonian is then
$H=H_0+H_{hybrid}+H_{int}$, with $H_0$ containing the primitive site
energies of $d$-, $p$-, and $f$-electrons denoted by
$\varepsilon^{(\alpha)}_d$, $\varepsilon^{(\mu)}_p$, and
$\varepsilon^{(m)}_f$, respectively.

It is expected that $U_p$ is small compared to the other Coulomb
interactions. We will therefore set $U_p = 0$, in which case the
$p$-orbitals can be readily integrated out. The obtained effective
Hamiltonian ${\tilde H}$ takes the form
\begin{eqnarray}
{\tilde H}=H_{0}+H_{d}+H_{f}+H_{df}+H_{int,d}+H_{int,f}\;.
\end{eqnarray}
Here $H_{d}=\sum_{{\vec
r}}V^{(\alpha\alpha')}_{d}[D^{(\alpha)\dagger}_{\sigma}({\vec
r})D^{(\alpha')}_{\sigma}({\vec r})+h.c.]$, $H_{f}=\sum_{{\vec
r}}V^{(m m')}_{f}[F^{(m)\dagger}_{\sigma}({\vec r}_f,\eta
z_f)F^{(m')}_{\sigma}({\vec r}_f,\eta z_f)+\text{h.c.}]$, and
$H_{df}=\sum_{{\vec r}}V^{(\alpha
m)}_{df}[D^{(\alpha)\dagger}_{\sigma}({\vec
r})F^{(m)}_{\sigma}({\vec r}_f,\eta z_f)+\text{h.c.}]$, with
$V^{(\alpha\alpha')}_{d}=-\sum_{\mu}
V^{(\mu,\alpha)}_{pd}V^{(\mu,\alpha')}_{pd}/\varepsilon^{(\mu)}_{p}$,
$V^{(m m')}_{f}=-\sum_{\mu} V^{(\mu,m)}_{pf}V^{(\mu,
m')}_{pf}/\varepsilon^{(\mu)}_{p}$, $V^{(\alpha
m)}_{df}=-\sum_{\mu}V^{(\mu,\alpha)}_{pd}
V^{(\mu,m)}_{pf}/\varepsilon^{(\mu)}_p$. In the momentum ${\bf
K}$-space (in the reduced Brillouin zone corresponding to two
Fe-atoms in the conventional cell with lattice constant $a=\sqrt
2$), $H_{d}= \sum_{{\bf
K}}V^{(\alpha\alpha')}_{d}g^{(\eta\eta')}_{d} ({\bf
K})d^{(\alpha)\dagger}_{\eta{\bf K}\sigma}d^{(\alpha')}_{\eta'{\bf
K}\sigma}$, $H_{f}=\sum_{{\bf K}}V^{(m m')}_{f}g_{f} ({\bf
K})f^{(m)\dagger}_{\eta{\bf K}\sigma}f^{(m')}_{\eta{\bf K}\sigma}$,
and $H_{df}=\sum_{{\bf K}}V^{(\alpha m)}_{df}g^{(\eta\eta')}_{df}
({\bf K})[d^{(\alpha)\dagger}_{\eta{\bf K}\sigma}f^{(m)}_{\eta'{\bf
K}\sigma}+\text{h.c.}]$, where $d^{(\alpha)}_{\eta{\bf K}\sigma}$
and $f^{(m)}_{\eta{\bf K}\sigma}$ are the Fourier transform of $d$-
and $f$- electron operators in  the sublattices $\eta=A$ or $B$,
respectively. The ${\bf K}$-dependence of the dispersions and
$d$-$f$ hybridization is only encoded in the form factors, given by
$g^{(AA)}_d({\bf K})=g^{(BB)}_d({\bf K})=4+2\cos (K_xa) \cos
(K_ya)$, $g^{(AB)}_d({\bf K})=g^{(BA)}_d({\bf K})=8\cos (K_xa/2)
\cos (K_ya/2)$, $g_{f}({\bf K})=16\cos^2 (K_xa/2)\cos^2 (K_ya/2)$,
$g^{(AA)}_{df}({\bf K})=g^{(BB)}_{df}({\bf K})=8\cos^2 (K_xa/2) \cos
(K_ya/2)$, $g^{(BA)}_{df}({\bf K})=g^{(BA)}_{df}({\bf K})=8\cos
(K_xa/2)\cos^2 (K_ya/2)$.

{\it The $d$-electron correlations.} For moderate large $U_d$, we
may start from the strong coupling limit yielding the frustrated
$J_1$-$J_2$ Heisenberg model for the
$d$-electrons~\cite{Si:08,Yildirim:08,Ma:08}. The itinerancy of the
$d$-electrons will further reduce the ordered moments and eventually
lead to a paramagnetic phase~\cite{DaiSiZhu:08}. In fact, both the
weak- and strong-coupling limits suggest that the staggered
magnetization $M_d=\sum_{\alpha}M^{(\alpha)}_d=-(1/N)\sum_{\bf
K}\langle \{d^{(\alpha)\dagger}_{\eta{\bf K}\uparrow}
d^{(\alpha)}_{\eta{\bf K+Q}\uparrow}-d^{(\alpha)\dagger}_{\eta{\bf
K}\downarrow} d^{(\alpha)}_{\eta{\bf K+Q}\downarrow}\}\rangle $ is a
dominating order parameter with $Q=(\pi,\pi)$ and $N$ being the
number of $\mathbf{K}$ points in the reduced Brillouin zone. For the
purpose of demonstrating the effect of $d$-electron order on the
Kondo effect, we treat $M^{(\alpha)}_d$ as the mean field parameters
and approximate $H_{int,d}$ by $J_d\sum_{{\bf K}} \sigma
M^{(\alpha)}_d [d^{(\alpha)\dagger}_{\eta{\bf K}\sigma}
d^{(\alpha)}_{\eta{\bf K+Q}\sigma}+\text{h.c.}]$, with $J_d$ being
the effective coupling strength. The AF ordering gap,
$\Delta_{AF}^{(\alpha)}=J_d M_d^{(\alpha)}$, is sizable for FeAs but
vanishes for FeP.

{\it Kondo effect vs. d-electron ordering.} In order to understand
the competition between the Kondo effect and $d$-electron AF order,
we first neglect the $f$-electron ordering. We are then led
to consider
\begin{eqnarray}
\label{ALM}
H_{ALM}&=&\sum_{\bf
K}[\varepsilon^{(\alpha)}_{d}\delta_{\alpha\alpha'}\delta_{\eta\eta'}
+V^{(\alpha\alpha')}_d g^{(\eta\eta')}_d({\bf
K})]d^{(\alpha)\dagger}_{\eta{\bf K}\sigma}d^{(\alpha')}_{\eta'{\bf
K}\sigma}\nonumber\\
&+&\sum_{\bf K}[\varepsilon^{(m)}_{f}\delta_{m m'}+V^{(m m')}_f
g_f({\bf K})] {f}^{(m)\dagger}_{\eta{\bf
K}\sigma}{f}^{(m')}_{\eta{\bf
K}\sigma}\nonumber\\
&+&\sum_{{\bf K}}[V^{(\alpha m)}_{df} g^{(\eta\eta')}_{df}({\bf
K})d^{(\alpha)\dagger}_{\eta{\bf K}\sigma}{f}^{(m)}_{\eta'{\bf
K}\sigma}+\text{h.c.}]\nonumber\\
&+&\sum_{{\bf K}}[\sigma \Delta_{AF}^{(\alpha)}
d^{(\alpha)\dagger}_{\eta{\bf K}\sigma} d^{(\alpha)}_{\eta{\bf
K+Q}\sigma}+\text{h.c.}]\nonumber\\
&+&U_f\sum_{{\vec r}_f} n^{(m)}_{f,\uparrow}({\vec r}_f,\eta z_f)
n^{(m)}_{f,\downarrow}({\vec r}_f,\eta z_f)\;.
\end{eqnarray}
In the absence of $d$-electron ordering, Eq.(4) is the ALM with weak
$f$-electron dispersion and momentum-dependent hybridization. (The
effect of momentum-dependent hybridization on the Kondo effect has
recently been studied in other contexts~\cite{Ghaemi:08,Weber:08}.)
For sufficiently large $U_f$, and with the $f$-levels being well
below Fermi energy, we are in the Kondo limit.

To concretely demonstrate how the $d$-electron AF order influences
the Kondo effect, we consider the resulting Kondo lattice model with
a single $f$-electron channel and two $d$-electron bands. In the
slave-boson representation, this becomes
\begin{eqnarray}
H_{KLM} &=& \sum_{\mathbf{k}}\epsilon_{d}^{(\alpha\alpha^{\prime})}
(\mathbf{k}) {d_{\mathbf{k}\sigma}^{(\alpha)}}^{\dagger}
d_{\mathbf{k}\sigma}^{(\alpha^{\prime})}
 + \lambda\biggl{(}\frac{1}{N_{L}}\sum_{\mathbf{k}}
f_{\mathbf{k}\sigma}^{\dagger}f_{\mathbf{k}\sigma} - 1\biggr{)}
\nonumber \\
&&+ \sum_{\mathbf{k}} [\sigma \Delta_{AF}
{d_{\mathbf{k}\sigma}^{(\alpha)}}^{\dagger}
d_{\mathbf{k}+\mathbf{Q},\sigma}^{(\alpha)}
+ \text{H.c.}] \nonumber \\
&& - \frac{1}{2} J_{K} \sum_{\mathbf{k}}
V_{df}(\mathbf{k})b_{\alpha} [f_{\mathbf{k}\sigma}^{\dagger}
d_{\mathbf{k}\sigma}^{(\alpha)} + \text{H.c.}].
\end{eqnarray}
Here, the Lagrange multiplier $\lambda$ enforces the single
occupancy of $f$-electrons. The mean-field parameter $b_{\alpha}=
\langle f_{\mathbf{k}\sigma}^{\dagger}
d_{\mathbf{k}\sigma}^{(\alpha)} \rangle/2$ describes the Kondo
screening and sets the Kondo scale, $T_K \propto b^2$. The
anisotropic hybridization form factor $V_{df}(\mathbf{k}) = 4 \cos
k_x/2 \cos k_y/2$. The energy dispersion for $d$-electrons are taken
to be~\cite{SRaghu:08}:
\begin{eqnarray*}
\epsilon^{(1)}(\mathbf{k}) &=& -2t_1 \cos k_x - 2t_2 \cos k_y - 4t_3 \cos k_x \cos k_y\;, \\
\epsilon^{(2)}(\mathbf{k}) &=& -2t_2 \cos k_x - 2t_1 \cos k_y - 4t_3 \cos k_x \cos k_y\;, \\
\epsilon^{(12)}(\mathbf{k}) &=& \epsilon^{(21)}(\mathbf{k}) = -4t_4 \sin k_x \sin k_y \;,
\end{eqnarray*}
with $t_1=-1$, $t_2=1.3$, $t_3=t_4=-0.85$.  In our numerical study,
we choose $J_K=0.04$, temperature $T=10^{-10}\vert t_1\vert$, and
the lattice size $N_{L} = 3200\times  3200$. When we vary the AF
order parameter, the chemical potential is adjusted such that the
$d$-electrons are fixed at the half-filling $n_d=2.0$.
\begin{figure}[t]
\epsfxsize=6.5cm \centerline{\epsffile{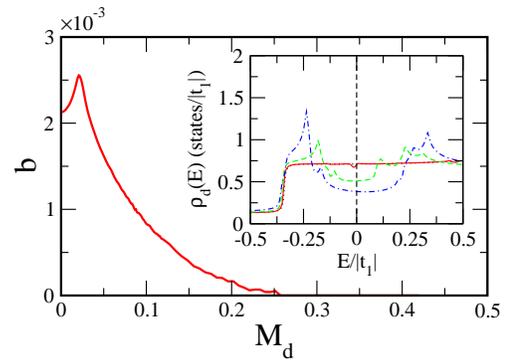}}
\caption{(Color online) Mean-field Kondo parameter $b$ as a function
of the $\mathbf{Q}=(\pi,0)$ (in the notation of the
one-Fe Brillouin zone) staggered magnetization $M_d$.
The Kondo temperature $T_K \propto b^2 $.
$M_d$ is measured in $\mu_B/$Fe.
The inset shows the $d$-electron density of states
for $M_d=0$ (dotted-black), 0.021 (black-red), 0.165
(dashed-green), 0.296 (dash-dotted-blue). }
\label{fig:Kondo-coupling}
\end{figure}

Fig.~\ref{fig:Kondo-coupling} shows that the
$d$-electron AF order rapidly suppresses the Kondo scale.
This suppression is closely related to the depression
of the $d$-electron density of states (DOS) in
the collinear AF state of undoped iron arsenides (see the inset of
Fig.~\ref{fig:Kondo-coupling}). The feature of low energy DOS is
sensitive to the degree of nesting and the DOS minimum is not
necessarily located precisely at the Fermi energy (see, {\it e.g.},
the case of $M_d=0.021$ in Fig.~\ref{fig:Kondo-coupling});
the latter explains the effective Kondo scale first rising
and then dropping with the AF order. Furthermore, the incomplete
nesting of the Fermi surface
keeps the depressed DOS finite (unlike, say, in the superconducting
state) at the Fermi energy such that the $T=0$ ground state has the
$f$-moment always Kondo screened on the lattice.

We should stress that, for the purpose of a semi-quantitative
assessment of the proposed mechanism, we have considered the upper
limit for the Kondo scale in the AF state: we have coupled the
$f$-moments to only the quasiparticles of the $d$-electron AF state
and have also neglected the $f$-moment ordering; moreover, a genuine
$f$-electron quantum phase transition will be induced by breaking
the Kondo screening upon the inclusion of the standard RKKY-Kondo
competition~\cite{QSi01,TSenthil04,IPaul07}. We can therefore infer
that the mechanism proposed here provides a viable basis to
understand the distinct $f$-electron heavy fermion behaviors in
CeOFeP ($M_{d}\approx 0$) and CeOFeAs ($M_{d}\approx
0.8$\cite{Zhao:08}). Our results also set the stage for
understanding the evolution of the heavy fermion behavior in the
CeOFeAs$_{1-x}$P$_{x}$ series. In general, there will be two
magnetic quantum critical points $x_{c_1}$ and $x_{c_2}$, associated
with the $d$- and $f$-electrons, respectively. The RKKY interaction
would then dominate in the intermediate region of $x$, leading
likely to a ferromagnetic order before the heavy fermion state is
approached.

\begin{figure}[t] \epsfxsize=8.0cm
\centerline{\epsffile{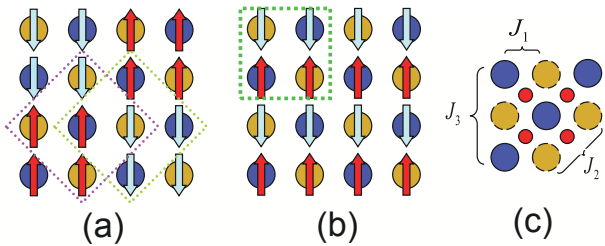}} \caption[]{(Color online) The
would-be ordering patterns of the $f$-electrons due to the
superexchange interactions via (a) the $R$-$X$-$R$ process alone or
(b) the  $R$-O-$R$ process alone. (c) illustrates the combined
exchanges, viewed as $J_1$-$J_2$-$J_3$ interactions of an {\it
effective} square lattice within an $R$O layer, which are expected
to turn the orders of (a) and (b) into a helical one. The blue and
brown circles label the rare-earth sites in the same way as in
Fig.~\ref{fig:lattice}. } \label{fig:f-ordering}
\end{figure}

{\it Magnetic frustration of the $f$-electrons.} We now turn to the
exchange interactions among the $f$-moments. Consider first the
superexchange interaction, which can be derived by integrating out
the virtual valence fluctuations of the $f$-electrons. From
Eq.~(\ref{ALM}), we end up with ${\tilde H}_{f}=\sum_{{\vec
r}}J^{(m,m')}_f {\vec S}^{(m)}_F({\vec r}_f,\eta z_f)\cdot {\vec
S}^{(m')}_F({\vec r}_f,\eta z_f)$, where ${\vec S}^{(m)}_F({\vec
r},\eta z_f)=\sum_{\square}{\vec S}^{(m)}_f ({\vec r}_f,\eta z_f)$
are summations of $f$-electron spins in the corresponding plaquettes
associated with ${\vec r}$, and $J^{(m,m')}_f\approx
2[V^{(mm')}_{f}]^2
(\frac{1}{U_f+\varepsilon^{(m)}_f}-\frac{1}{\varepsilon^{(m')}_f})$.
This is the superexchange interaction associated with the
$R$-$X$-$R$ path, which does not mix the odd and even sublattices of
the $f$-sites in a single $R$O layer (see
Fig.~\ref{fig:f-ordering}(a)). There will also be a superexchange
interaction from the $R$-O-$R$ path, due to the hybridization
between the 4$f$-orbitals of $X$-atoms and the 2$p$-orbitals of
O-atoms; this superexchange mixes the odd and even sublattices (see
Fig.~\ref{fig:f-ordering}(b)). In the notations of an effective
square lattice of the $f$-sites ({\it c.f.}
Fig.~\ref{fig:f-ordering}(c)) the $R$-O-$R$ path gives rise to the
nearest-neighbor (n.n.) interaction $J^{(O)}_1$ and the
next-nearest-neighbor (n.n.n.) $J^{(O)}_2$, while the $R$-$X$-$R$
path yields the n.n.n. $J^{(X)}_2$ and the third-nearest-neighbor
(n.n.n.n.) $J^{(X)}_3$. (Note that $J^{(X)}_2$ and $J^{(X)}_3$
correspond to the n.n. and n.n.n. interactions in the odd/even
sublattices separately.) The resulting $f$-electron spin Hamiltonian
becomes a $J_1$-$J_2$-$J_3$ Heisenberg model
(Fig.~\ref{fig:f-ordering}(c))
 \begin{eqnarray}
{\cal H}_f=\displaystyle\{\sum_{n.n.}J_1+ \sum_{n.n.n.}J_2+
\sum_{n.n.n.n.}J_3\}{\vec S}_i\cdot {\vec S}_j~,
 \end{eqnarray}
where $J_1=J^{(O)}_1$, $J_2=J^{(O)}_2+J^{(X)}_2$, and
$J_3=J^{(X)}_3$.

In this way, the $f$-moments of CeOFeAs provides a realization of a
geometrically frustrated quantum magnetic system in two dimensions.
Quantum frustrated magnets have been the subject of theoretical
studies for a long time, and continue to attract extensive interest
~\cite{Diep}. However, suitable materials with spin-$1/2$ are rare.
In this context, it will be very important to clarify the magnetic
behavior of the $f$-moments in CeOFeAs and related arsenides.

The ${\vec S}_F\cdot {\vec S}_F$ form given earlier corresponds to
$J_3^{(X)}/J_2^{(X)}$ being equal to $1/2$, and further bond-angle
considerations imply that $J_3^{(X)}/J_2^{(X)}$ will be larger than
$1/2$ but still not far away from it. Similar considerations would
suggest that $J_2^{(O)}/J_1^{(O)} \approx 1/2$. We will therefore
expect $J_2>J_1/2$ and a sizable $J_3/J_1$. In this range, the
N\'{e}el and collinear orderings are excluded. Instead, an
incommensurate helical phase with the ordering vector $(q,\pi)$ or
$(q,q)$ is the most likely ground state, where $\cos
q=\frac{2J_2-J_1}{4J_3}$~\cite{Moreo:90}. Neutron scattering and
muon spin relaxation experiments in polycrystal CeOFeAs appear to
have seen a helical $f$-electron ordering~\cite{Zhao:08,Maeter:08}.

In the $d$-electron paramagnetic regime, there will also be an RKKY
interaction. The latter is expected to be ferromagnetic given the
relatively small size of the Fermi surfaces, and this is consistent
with the enhanced ferromagnetic fluctuations of the heavy fermion
state observed in CeOFeP~\cite{Bruning:08}.
%Interestingly, the same
%mechanism can also explain the distinction between two homologous
%compounds, CeORuP and CeOOsP. While the former shows the heavy
%fermion behavior with ferromagnetic fluctuations, the latter is a AF
%metal without heavy fermion behavior~\cite{Krellner:08}.
Still, the
frustrating $J_1$-$J_2$-$J_3$ superexchange interactions will
continue to operate, helping to suppress the tendency for AF
ordering.

{\it Discussion and summary.} A number of other consequences of the
$p$-$f$ hybridization are relevant to the iron-pnictides phase
diagram. First, in the heavy fermion phase, the momentum-dependence
of the induced $d$-$f$ hybridization will generally smear the
hybridization gap (which has nodal lines along ${\bf K}_x=\pm\pi/a$
and ${\bf K}_y=\pm\pi/a$), and this could be visible in the
optical-conductivity spectrum.
% Second, the Kondo coupling may
%be manifested even in the superconducting case, as it may affect
%the $d$-electron pairing at the lowest temperatures. Third,
Second, the induced $d$-$f$ hybridization depends on the Fe-$X$ and
$X$-$R$ distances. Increasing pressure along the $c$-axis will
decrease the distances and increase the hybridizations, and
eventually enhance $T_K$ \cite{Georges:08}. Finally,
%Second,
in light of the fact that
the $f$-electron ordering is further suppressed by the competing
$J_1$-$J_2$-$J_3$ interactions, the transition or crossover from
the superconducting to the heavy fermion phases may take place
at sufficiently high pressures in the carrier-doped superconducting
materials.

In summary, we have considered a mechanism for weakening the Kondo
screening effect through the antiferromagnetic order of the
conduction electrons, and implemented it in an extended Anderson
lattice Hamiltonian. For the iron pnictides, our mechanism is
semi-quantitatively viable to explain the observed existence/absence
of heavy fermion behavior in CeOFeP and CeOFeAs, respectively. More
broadly, our mechanism goes beyond the standard picture of heavy
fermion physics, {\it viz.} the RKKY and Kondo competition, and can
therefore shed new light on the phase diagram of heavy fermion
systems in general. Finally, we have proposed that the $f$-electrons
in the parent iron arsenides represent a rare model system for
quantum frustrated magnetism in two dimensions.

We thank E. Abrahams, M. Aronson, G. H. Cao, X. H. Chen, X. Dai, C.
Geibel, N. L. Wang, T. Xiang, Z. A. Xu, and H. Q. Yuan for useful
discussions, and the U.S. DOE CINT at LANL for computational
support. This work was supported by the NSF of China, the 973
Program, and the PCSIRT (IRT-0754) of Education Ministry of China
(J.D.), the NSF Grant No. DMR-0706625 and the Robert A. Welch
Foundation (Q.S.), and by U.S. DOE at LANL under Contract No.
DE-AC52-06NA25396 (J.-X.Z.).

\end{document}